\documentclass[usenatbib,useAMS]{mn2e}
\usepackage{graphicx}
\usepackage{amsmath}
\usepackage{amssymb}
\usepackage{lscape}
\usepackage{color}
\usepackage{soul}

\def\grad{{\rm grad}\,}

\def\be{\begin{equation}}
\def\ee{\end{equation}}
\def\bea{\begin{eqnarray}}
\def\eea{\end{eqnarray}}

\def\lapprox{\mathrel{\hbox{\rlap{\hbox{\lower4pt\hbox{$\sim$}}}\hbox{$<$}}}}
\def\gapprox{\mathrel{\hbox{\rlap{\hbox{\lower3pt\hbox{$\sim$}}}\hbox{$>$}}}}

\title[Variations of $P_2$ in subpulse drifting pulsars]{Variations of $P_2$ in subpulse drifting pulsars}
\author[R. Yuen et. al.]
{R. Yuen$^{1,2,3}$\thanks{E-mail: ryuen@xao.ac.cn}, D. B. Melrose$^3$, M. A. Samsuddin$^4$, Z. Y. Tu$^1$ and X. H. Han$^1$\\
$^1$Xinjiang Astronomical Observatory, Chinese Academy of Sciences, 150 Science 1-Street, Urumqi, Xinjiang, 830011, China\\
$^2$Key Laboratory of Radio Astronomy, Chinese Academy of Sciences\\
$^3$SIfA, School of Physics, University of Sydney, Sydney, NSW 2006, Australia\\
$^4$University of Malaya, Malaysia}
\date{%Accepted \hspace{2cm}
          --- Received in original form}
\pubyear{2012}

\begin{document}

\maketitle
\begin{abstract}
We develop a model for subpulse separation period, $P_2$, taking into account both the apparent motion of the visible point as a function of pulsar phase, $\psi$, and the possibility of abrupt jumps between different rotation states in non-corotating pulsar magnetospheres. We identify three frequencies: (i) the spin frequency of the star, (ii) the drift frequency of the magnetospheric plasma in the source region, and (iii) the angular frequency of the visible point around its trajectory. We show how the last of these, which is neglected in traditional models by implicitly assuming the line of sight through the center of the star, affects the interpretation of $P_2$. We attribute the subpulse structure to emission from $m$ anti-nodes distributed uniformly in azimuthal angle about the magnetic axis. We show that variations of $P_2$ as a function of rotational phase or observing frequency arise naturally when the motion of the visible point is taken into account. We discuss possible application of our model in signifying overall field-line distortion at the emitting region. Abrupt changes in $P_2$ can occur during state switching in the magnetosphere. We demonstrate that the unique value of $P_2$ in each rotation state can be used, in principle, to relate the rotation state of the magnetospheres to subpulse drifting.
\end{abstract}

\section{Introduction}

Drifting subpulses appear as a modulation of pulsar radio emission in which subpulses drift through the (integrated) pulse window. The separation in time between two adjacent subpulses is denoted  by $P_2$ \citep{SSP+70, Backer73}, which is shorter than $P_1=2\pi/\omega_*$ \citep{RudermanSutherland1975, MAZ14}, where $\omega_*$ is the rotation frequency of the star. In the carousel model \citep{DR99} subpulses are interpreted as emission from plasma columns distributed uniformly around the magnetic pole and rotating at $\omega_{\rm dr}\ne\omega_*$. One possible interpretation of the ``plasma columns'' is in terms of anti-nodes\footnote[1]{Assuming either anti-nodes or nodes does not affect our calculations.} of a wave-like structure, generated by a large-scale instability in the magnetosphere \citep{Arons1981, Jones1984, KMM91a}. The diocotron instability seems a plausible generation mechanism \citep{FKK06, LT12}. Assuming a wave at a specific spherical harmonic grows preferentially \citep{RudermanSutherland1975, CR04, GMM+05}, the resulting pattern is proportional to $Y_l^m(\theta_b,\phi_b)$, where $l$ and $m$ are integers and $\theta_b$ and $\phi_b$ are polar and azimuthal angles, respectively,  relative to the magnetic axis. Such a pattern, $\propto\exp(im\phi_b)$, corresponds to $m$ anti-nodes. The wave structure may be propagating relative to the magnetospheric plasma, but for simplicity  the structures are usually assumed to be comoving with the plasma. A detailed model for drifting subpulses is needed to use the data to determine the actual motion of magnetospheric plasma, and hence to provide information on the electrodynamical processes that drive this motion. A model for drifting subpulses also needs to accommodate other observational features, such as curved driftbands and abrupt changes in $\omega_{\rm dr}$, so that these may be used to infer further details regarding the magnetosphere.

Curved driftbands  are common  in radio pulsars  that exhibit drifting subpulses \citep{ES03, SMK05, SSW09, HSW+13}. The ``curvature'' describes how $P_2$ varies across the pulse window. 
%It requires that the model of a uniformly rotating distribution of $m$ equally spaced anti-nodes be modified in some way. 
There is evidence, from observations at both radio and $\gamma$-ray frequencies, that a pulsar magnetosphere may switch between different rotation states, affecting the subpulse drift pattern \citep{SLS00, LHK+10, SMK05, KLO+06, ABB+13, KSJ13}. By ``rotation state'' we mean a specific value of $\omega_{\rm dr}$ (in the emission region), and the evidence implies that a pulsar may jump abruptly between two or more different rotation states. A model for non-corotational motion is needed to describe the rotation states between which such jumps occur.

In this paper we consider a model for drifting subpulses that combines a quantitative model for the non-corotational motion (of the magnetosphere of an oblique rotator) and an effect that has conventionally been ignored: the apparent motion of the visible point. Our objectives are to develop a quantitative model for the non-corotational motion and apply it to observed changes associated with jumps between rotation states, and to explore the implications of the motion of the visible point on subpulses, specifically, the observed variations of $P_2$ in drifting subpulses.

Our quantitative model \citep{MY14a} is based on an interpolation of the plasma motions between two limiting cases. One limit corresponds to the vacuum-dipole model, in which there is no plasma, and the (inductive) electric field, ${\bf E}_{\rm ind}$, is determined by the rotating magnetic dipole. In this case, any test charge has an electric drift across the magnetic field lines, at a velocity ${\bf v}_{\rm ind}$ say. We interpret ${\bf v}_{\rm ind}$ as the plasma flow velocity in the limit where the plasma density is arbitrarily small, such that there are too few charges to provide the (additional) potential field implied by corotation. The other limit corresponds to corotation, in which case the plasma velocity across the magnetic field lines is the perpendicular component of the corotation velocity, ${\bf v}_{\rm cor\perp}$ say, which is the electric drift in the corotation electric field, ${\bf E}_{\rm cor}$. (The plasma motion along the magnetic field lines is unconstrained by the electrodynamics.) We consider a one-parameter class of rotation states described by the parameter $y$, such that the actual plasma motion across the field lines is $y{\bf v}_{\rm ind}+(1-y){\bf v}_{\rm cor\perp}$. By construction this model satisfies the requirement that the flow pattern is periodic, with period $P_1$, although the motion of an individual plasma blob is not periodic (except %to
for $y=0$). In this model, an abrupt change in rotation state is parametrized by a change in $y$.  

The apparent motion of the visible point follows from the long-accepted assumption that pulsar emission is confined to a tiny cone about the tangent to the local magnetic field line \citep{RC69_RVM}. This implies that emission is visible only by an observer whose line of sight is parallel to the local magnetic field line. The line of sight to this ``visible point'' varies in the transverse direction with the phase of an oblique rotator, so that the visible point moves around a trajectory at an apparent velocity that is determined purely by geometry and the assumed magnetic field structure \citep{Gangadhara04, YM14}. (Existing models for drifting subpulses are based on the implicit, but incorrect, assumption that the line of sight is a fixed line, e.g., through the center of the star.)

In Section \ref{sect:visibility} we discuss the apparent motion of the visible point. In Section \ref{sect:3Periods} we summarize three motions that can contribute to the observed $P_2$. Variations in $P_2$ are discussed in Section \ref{sect:P2Variations} in three different, but related, contexts, and a test case using PSR B0031-07 for our model is considered. We discuss the results and summarize our conclusions in Section \ref{sect:Conclusions}.

\section{Visibility of emitting anti-nodes}
\label{sect:visibility}

In this section, we summarize the viewing geometry \citep{YM14} %for use in describing the visible anti-nodes.
assumed in this paper.

\subsection{Visible points in a magnetosphere}
\label{sect:VisibleEmissionPoints}

The angular position of the visible point on a sphere of radius $r$ is independent of $r$ in a dipolar magnetic field. Let this position be described by polar and azimuthal angles, $\theta = \theta_V (\zeta,\alpha;\psi)$ and $\phi = \phi_V (\zeta,\alpha;\psi)$, in the observer's frame, or  by $(\theta_{bV}, \phi_{bV})$, where the subscript $b$ represents quantities in the magnetic frame. Here $\zeta$ and $\alpha$ correspond to the viewing angle between the rotation axis and the line of sight and the obliquity angle between the rotation and magnetic axes, respectively, and $\psi$ is the rotational phase of the pulsar. The transformation matrix between the two coordinate systems is given in Appendix \ref{sec:CoordTansf}. The visible point in the magnetic frame is then given by \citep{Gangadhara04, YM14}
\begin{eqnarray}
\cos 2\theta_{b{\rm V}} = \frac{1}{3} (\cos\theta_b \sqrt{8+\cos^2\theta_b} - \sin^2\theta_b), & \nonumber \\
\tan\phi_{b{\rm V}} = \frac{\sin\zeta\sin\psi}{\sin\alpha\cos\zeta-\cos\alpha\sin\zeta\cos\psi},
\label{EmissionPtPhiB}
\end{eqnarray}
with $\cos\theta_b = \cos\alpha\cos\theta + \sin\alpha\sin\theta\cos(\phi-\psi)$. For given $\zeta,\alpha$ the visible point traces out a closed path (on the sphere of radius $r$) in one pulsar rotation, which we refer to as the trajectory of the visible point. 

The visibility of pulsar emission is dictated by the trajectory of the visible point whose geometry depends strongly on $\zeta,\alpha$ (see Figure \ref{fig-TrajMeetNodes}). The value of $\theta_V$ is maximum at $\psi =0$ and minimum at $\psi =180^\circ$.  The traditional approach, in which the line of sight is assumed fixed,  is a valid approximation only in the limiting cases of $\zeta\cong 0$. We ignore the special case $\zeta =0$ in the following discussion.

Assuming that pulsar emission comes only from open-field regions restricts the source location to radii greater than a minimum visible height, $r_{\rm min}$. For given $\zeta,\alpha$ one has \citep{YM14}
\begin{equation} \label{eq:MinScrHeight}
r_{\rm min} = \frac{r_L  \sin^2 \theta_{b{\rm V}}}{\sin^2 \theta_{bL}(\phi_{b{\rm V}})\, \sin\theta_L(\phi_{b{\rm V}})},
\end{equation}
where $\theta_L$ is the angle measured from the rotation axis to the point where the last closed field line is tangent to the light cylinder. Emission is visible only while the trajectory of the visible point is inside the open-field region, and the maximum width of the pulse window is the range of $\psi$ over which this condition is satisfied. This width increases with increasing $r-r_{\rm min}$ and reduces to a point for $r=r_{\rm min}$. The pulse window is centered on $\psi=0$, where the rotation axis, magnetic axis and the line of sight are coplanar. The pulse width increases as the impact parameter, defined by $\beta = \zeta - \alpha$, decreases. 

\subsection{Model for visible emission}
\label{sect:ModelVisibleEmission}

A unique emission location in the magnetosphere may be defined by assuming that observable emission is only from the last closed field lines. In this model, the radial distance from the center of the star to any points on the field line at $\phi_b$, which is a function of $\psi$ (see Appendix A), can be determined using Equation (2) with polar angles given by the trajectory of the visible point and known field line constant. Observable emission in the magnetosphere is therefore specified by $(r_{\rm min}, \theta_V, \phi_V)$ on each last closed field lines around a pulsar. These parameters vary with pulsar phase, with $r_{\rm min}$ having its minimum at $\psi =0$ and its maximum  at $\psi = 180^\circ$. It is only in the aligned case, $\alpha = 0$, that $r_{\rm min}$ is independent of $\psi$.

\subsection{Emission height}

The height (or radius $r$) of the source of pulsar radio emission is poorly determined, and is usually estimated based on one of two methods: (i) a relativistic phase shift, which requires an asymmetric pulse profile with clearly defined core and cone components \citep{DRH04}, or (ii) geometry alone, with the source located on the last closed field lines \citep{KG03}. We assume model (ii). This implies that emission is visible from only one point in the magnetosphere for any given $\psi$. For example, if one assumes a unique radius-to-frequency mapping, then (ii) implies that emission at a given frequency is visible from only one height.

Assumption (ii) may be modified to relax the implied unique radius-to-frequency mapping. The emission is confined to a narrow forward cone (half-angle $1/\gamma$, where $\gamma$ is the Lorentz factor of the emitting particle), which is at a small angle to the magnetic field line due to aberration, so that there is some spread in emission about the direction tangent to the field line. More importantly, one expects emission from at least a small range of field lines inside the last closed field line, and this translates into emission over a corresponding small range of heights $r>r_{\rm min}$.

\subsection{Geometrical model for visible anti-nodes}

\begin{figure}
\begin{center}  % this centres figure in column
\includegraphics[width=1\columnwidth]{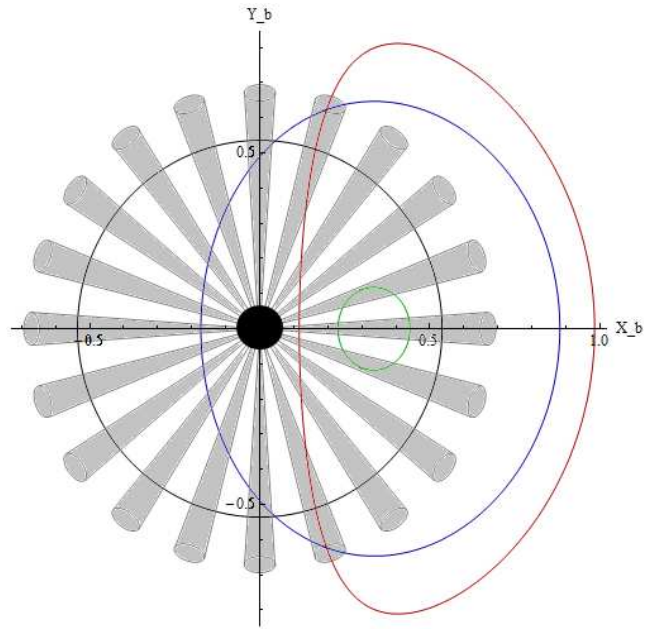}
\caption{Plots of the trajectory of the visible point (solid curves) are shown, projected onto the plane of the sky, for four choices of $(\zeta,\alpha)$: $(50^\circ,0)$ (black), $(10^\circ,30^\circ)$ (green), $(60^\circ,45^\circ)$ (blue) and $(70^\circ,80^\circ)$ (red). Also shown are radial spoke structure of aligned anti-nodes spaced equally around the star (black sold circle) in the magnetic frame centered on the magnetic pole. A trajectory starts at a point on the $x_b$-axis closest to the origin at $\psi =0^\circ$ then progresses in clockwise direction and intersects with the $x_b$-axis again at the farthest point from the origin reaching the highest height of the path at $\psi =180^\circ$, then returns to the starting point. Emission is visible in principle only from a spoke that is inside the relevant trajectory.}
\label{fig-TrajMeetNodes}
\end{center}
\end{figure}

We assume that there are $m$ emission sites, corresponding to anti-nodes, that are equally spaced in azimuth around the magnetic axis. The location of the anti-nodes is then independent of polar angle corresponding to alignment along the radial direction in a structure of radial spokes as illustrated in Figure \ref{fig-TrajMeetNodes}. The trajectory of the visible point intersects the spokes at different polar angles, $\theta_b$. The green, blue and red curves correspond to non-circular trajectories of the visible point in specific oblique models, and the dark black circle corresponds to a circular trajectory in a specific aligned case.  It is apparent from the figure that, for an oblique rotator, the trajectory is not concentric with the magnetic axis (e.g., red, blue and green), and need not enclose the magnetic axis (e.g., red and green). Moreover, it is only in the aligned case that all anti-nodes are potentially visible (when the restriction to emission only from within the polar cap is relaxed); for example, fewer than 50\% of the spokes are intersected by the red trajectory.  In traditional models \citep{ES03}, the number of anti-nodes is estimated from observation by implicitly assuming that all are visible in principle, as for an aligned rotator. This corresponds to the black curve, which is a circle centered on the magnetic pole.  We conclude, based on Figure \ref{fig-TrajMeetNodes}, that it is important to treat the geometry correctly, rather than identifying potentially observable subpulses from a single emission circle.

Note that the restriction to emission from within the polar cap is not taken into account in Figure \ref{fig-TrajMeetNodes}. The number of spokes intersected by a trajectory, and hence the number of visible anti-nodes, depends strongly on $r/r_{\rm min}>1$, with $r_{\rm min}$ given by Equation (\ref{eq:MinScrHeight}), and for a given $r/r_{\rm min}$ the polar-cap region defines an additional closed region in the figure \citep{YM14}. For the model discussed in Section \ref{sect:ModelVisibleEmission}, the trajectory cuts spokes at different $\theta_V,\phi_V$ and hence at different heights, implying that emission is from higher (lower) altitudes at the edges (center) of a profile \citep{GG01, KJ07}.

\section{Three frequencies relevant to subpulse drifting}
\label{sect:3Periods}

Three frequencies are relevant to a model for subpulse drifting: (i) the spin frequency of the star, $\omega_\star$, (ii) a frequency, $\bomega_V$, associated with the motion of the visible point, and (iii) the drift frequency of the plasma, $\bomega_{\rm dr}$. In this section, we discuss how these frequencies relate to $P_2$.

\subsection{Motion of the visible point}

The visible point moves at an angular velocity $\bomega_{\rm V}$ along the trajectory with components in the polar and azimuthal directions:
\begin{equation} \label{eq:VelEmPtComp}
\omega_{{\rm V}\theta} = \omega_\star \frac{\partial\theta(\alpha,\psi)}{\partial\psi}, \quad \omega_{{\rm V}\phi} = \omega_\star \frac{\partial\phi(\alpha,\psi)}{\partial\psi}, 
\end{equation}
where $\omega_\star = d\psi/dt$ is the angular speed of the star. The motion of the visible point is absent only for an aligned rotator, $\alpha = 0$. Another special case is for the line of sight along the rotation axis,  when the visible point moves only in the $\hat\bphi$ direction as the pulsar rotates. In the general case, the visible point sub-rotates when the magnetic axis is on the near side of the pulsar (near $\psi = 0$), and super-rotates when the magnetic axis is on the far side of the pulsar (near $\psi=180^\circ$). The average angular speed, $\langle \omega_V (\psi) \rangle = \omega_\star$, is equal to that of the star. For pulsars with a small pulse window, the observer sees only a small range of $\phi$ centered on $\phi= \psi= 0$ where $\omega_V$ is lowest.

\subsection{Plasma drift velocity}

The plasma drift velocity is modeled here as a linear combination of the electric drift velocities in the vacuum-dipole and  corotation models \citep{MY14a}. For an obliquely rotating pulsar the electric field is assumed to be of the form 
\begin{equation} \label{eq:E}
{\bf E} = (1-y{\bf b}\,{\bf b}\cdot){\bf E}_{\rm ind} -(1-y)\grad\Phi_{\rm cor},
\end{equation}
where ${\bf b}$ is the unit vector along the direction of the magnetic field, and where $-\grad\Phi_{\rm cor}$ is the potential electric field associated with the corotation charge density. (The corotation electric field is given by Equation (\ref{eq:E}) with $y=0$.) The electric drift velocity due to this electric field is
\begin{equation} \label{eq:driftVel}
{\bf v}_{\rm dr} = y{\bf v}_{\rm ind} + (1-y){\bf v}_{\rm cor\perp}.
\end{equation}
We write $\bomega_{\rm dr} = {\bf v}_{\rm dr}/r$, and note that only the non-radial components of ${\bf v}_{\rm dr}$ contribute to the apparent rotation frequency of the plasma.

\begin{figure}
\begin{center}  % this centres figure in column
\includegraphics[width=1\columnwidth]{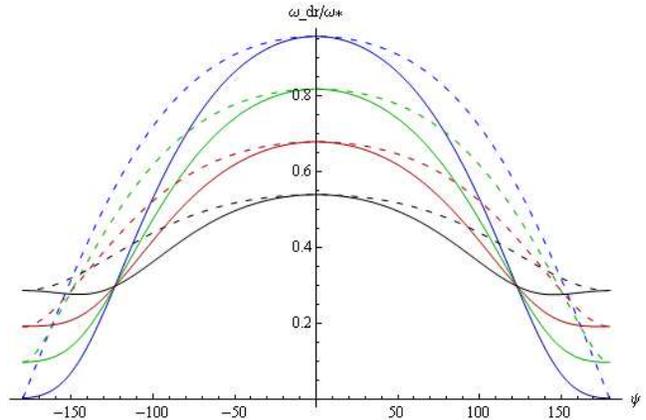}
\caption{The drift frequency, $\omega_{\rm dr}/\omega_\star$ (dotted), and its projection (solid) onto the trajectory of the visible point, are plotted against $\psi$ for $\zeta =70^\circ$ and $\alpha =80^\circ$ and $y =0$ (blue), 0.3 (green), 0.6 (red) and 0.9 (black).}
\label{fig-vDr_dis}
\end{center}
\end{figure}

The apparent observed ${\bomega}_{\rm dr}$ is the projection onto the trajectory of the visible point. Figure \ref{fig-vDr_dis} illustrates the change of $\omega_{\rm dr}/\omega_\star$ as a function of $\psi$ along the trajectory of the visible point (dotted) and the projected component (solid) for different $y$ values as given by Equation (\ref{eq:driftVel}). Except in limiting cases ($\zeta =0$, $\alpha=0,180^\circ$), which we ignore, the projected value is always lower than $\omega_{\rm dr}$. %The differences between the two increases as $\zeta$ increases. Generally, the two velocities converge for a small range of $\psi$ centered at $\psi =0/\pm 180^\circ$, where they both reach a maximum and minimum, respectively. The only non-zero component is along $\hat{\bphi}$ at $\psi =0$ and $180^\circ$ having the exact same expression from Equation (\ref{eq:driftVel}). The difference in value at the two rotational phases is due to the asymmetry of the trajectory around the magnetic axis. The two curves meet only at the two rotational phases imply that both are directed along the trajectory and the plasma drifts directly across the pulse window. 
We use the projected component for the following calculations.

\subsection{Motion of standing wave}

A purely magnetospheric model for drifting subpulses, as opposed to a model related to structures (e.g., ``hot spots'') on the surface of the star, requires that there be some quasi-stable structure in the source region. The assumption made here is that this is a standing wave at a specific spherical harmonic, with given $l$ and $m$, implying $m$ nodes and anti-nodes as a function of azimuthal angle. In general one expects such a wave to be propagating, at a velocity that depends on $l$ and $m$. For simplicity, we assume that the propagation velocity of the wave structure relative to the plasma is negligible. In a more general model the relative velocity of the wave structure to the plasma should be taken into account.

\subsection{Interval of subpulse separation}

Suppose one ignores both the motion of the visible point, $\omega_V\to0$, and the relative motion of the wave structure to the plasma. Then $m$ anti-nodes would pass through the (assumed fixed) line of sight in the time, $2\pi/\omega_{\rm dr}$, that it takes for the plasma to make a complete rotation. An observer would see consecutive anti-nodes separated by an interval
\begin{equation} \label{eq:P2t}
P_{2,S} (y) = \frac{2\pi}{m\omega_{\rm dr}}.
\end{equation}
This interval is modified when the motion of the visible point $(\omega_V \neq 0)$ is taken into account.  If the anti-nodes move faster (slower) than the visible point, emission from consecutive anti-nodes appears to arrive progressively earlier (later). The time interval (\ref{eq:P2t}) is replaced by
\begin{equation} \label{eq:P2InTime}
P_2(y) = \frac{2\pi}{m\omega_{\rm dr} -\omega_V}.
\end{equation}
Rewriting Equation (\ref{eq:P2InTime}) in units of $P_1$ gives
\begin{equation} \label{eq:P2InTime2}
\frac{P_2(y)}{P_1} = \big( mR_{\rm dr} -R_V \big)^{-1},
\end{equation}
where $R_{\rm dr} = R_{\rm dr}(y) = \omega_{\rm dr}/\omega_\star$ and $R_V = \omega_V/\omega_\star$. Note that the subpulse separation period is a function of $y$, potentially allowing $y$ to be determined from observations.

\subsection{Effect of motion of visible point}

We can make explicit the predicted effect of the motion of the visible point on $P_2$ by comparing Equations (\ref{eq:P2t}) and (\ref{eq:P2InTime}). The ratio of the value of $P_2$ when $\omega_V$ is artificially set to zero to that for $\omega_V\ne0$ is
\begin{equation} \label{eq:ratioParam}
\frac{P_2}{P_{2,S}} = R_d(y) = \bigg(1- \frac{\omega_V}{m\omega_{\rm dr}} \bigg)^{-1}.
\end{equation}
Equation (\ref{eq:ratioParam}) quantifies the discrepancy introduced, in traditional models, by assuming a fixed line of sight, rather than the line of sight that is always tangent to the field line in the source region \citep{YM14}. 

\begin{figure}
\begin{center}  % this centres figure in column
\includegraphics[width=1.\columnwidth]{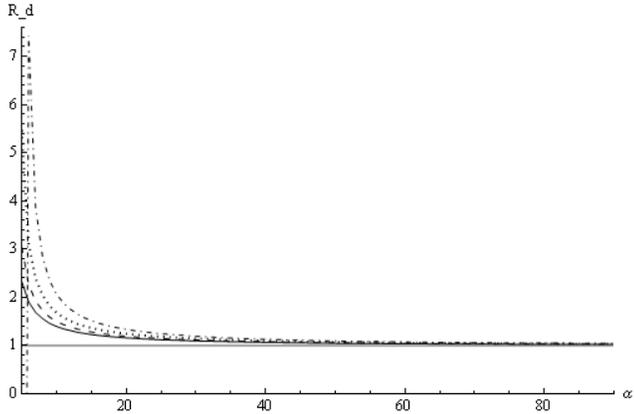}
\caption{Plot of the discrepancy $R_d$ at $\psi =0$ as a function of $\alpha =\zeta$ from $5^\circ - 90^\circ$ for $y =0$ (solid), 0.3 (dashed), 0.6 (dotted) and 0.9 (dot-dashed). The horizontal gray line {at $R_d = 1$ would correspond to no error being introduced in the predicted $P_2$ by neglecting the motion of the visible point.} The discrepancy is always greater than one and increases as either $\alpha$ decreases or $y$ increases. %For $y =0.9$, $R_d \approx -25$ at $\alpha =5^\circ$.
}
\label{fig-ExpansionFactor}
\end{center}
\end{figure}

Figure \ref{fig-ExpansionFactor} illustrates $R_d$ at $\psi =0$ for $\alpha = \zeta$ ranging from $5^\circ$ to $90^\circ$ for four different $y$ values. The discrepancy increases as $\alpha$ decreases and approaches unity at large $\alpha$. This implies that the traditional models underestimate $P_2$, giving a good approximations only for pulsars with $\alpha \geq 45^\circ$. Simulations using different values of $m$ give similar results although better agreement is obtained for $m \gtrsim 30$. 

\section{Variations of $P_2$}
\label{sect:P2Variations}

%Non-constant $P_2$ arises from the relative motion between $\omega_{\rm dr}$ and $\omega_V$ leading to longitudinal variations, frequency-dependent variations and abrupt changes in $P_2$ due to switching between different rotation states.

In this section we discuss several different effects that can cause a non-constant $P_2$.

\subsection{Changes with rotational phase}

\begin{figure}
\begin{center}  % this centres figure in column
\includegraphics[width=1.\columnwidth]{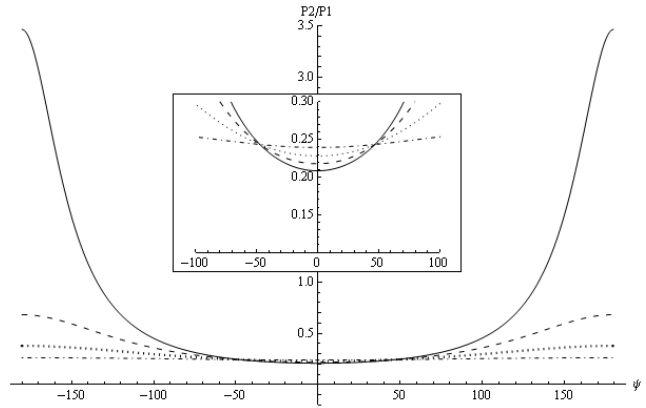}
\caption{Variations of $P_2$ in units of $P_1$ for $\zeta =10^\circ$ and $\alpha =30^\circ$ with $m=20$ for $y=0$ (solid), 0.3 (dashed), 0.6 (dotted) and 0.9 (dot-dashed) plotted against $\psi$. The inset magnifies the region for $-100^\circ \leq\psi\leq 100^\circ$ revealing minute changes in $P_2$ around $\psi =0$.}
\label{fig-P2_1}
\end{center}
\end{figure}

\begin{figure}
\begin{center}  % this centres figure in column
\includegraphics[width=1.\columnwidth]{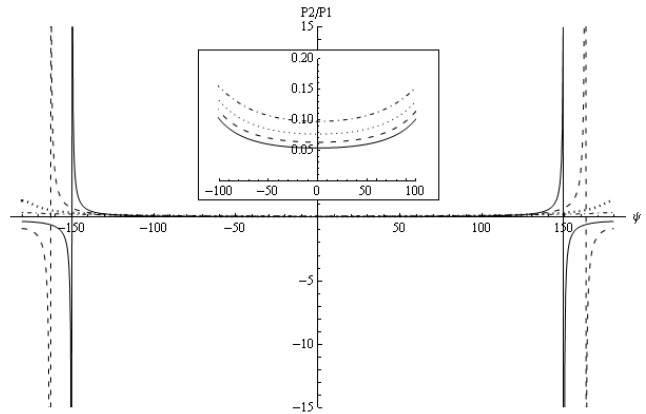}
\caption{Same as in Figure \ref{fig-P2_1} but for $\zeta =70^\circ$ and $\alpha =80^\circ$. $P_2$ changes sign at large $\psi$ for small $y$.}
\label{fig-P2_2}
\end{center}
\end{figure}

\begin{figure}
\begin{center}  % this centres figure in column
\includegraphics[width=1.\columnwidth]{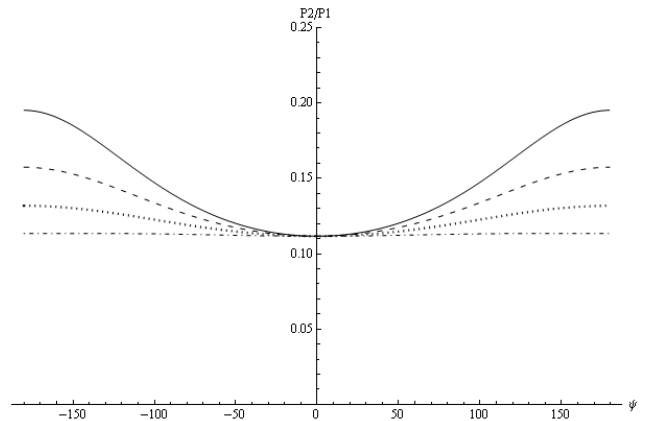}
\caption{Same as in Figure \ref{fig-P2_1} but for $\zeta =10^\circ$ and $\alpha =65^\circ$. $P_2$ for different $y$ is indistinguishable for $\Delta\psi \ll 50^\circ$ around $\psi =0$.}
\label{fig-P2_3}
\end{center}
\end{figure}

Simulations based on Equation (\ref{eq:P2InTime2}) show that the value of $P_2$ varies as a function of $\psi$ with the variability decreasing as $y$ increases. For large $y\to1$, corresponding to a near-vacuum model, $P_2$ is mostly constant throughout the whole rotational phase for small $\zeta$ as shown in Figures \ref{fig-P2_1} -- \ref{fig-P2_3}. The anti-node number is chosen to be $m = 20$ for simulations in this paper \citep{GMM+05, MR08}. The subpulse separation period, $P_2$, has its maximum at $\psi =180^\circ$,  on the far side of the magnetosphere relative to the line of sight, and its minimum at $\psi = 0$. The rate of change of $P_2$ with $\psi$ is smaller for larger $\alpha$ for a given $y$ with $\psi$ remaining $\lesssim 1\,$rad. It follows that for the pulsars with narrower pulse profiles, variations in $P_2$ are smaller.

For the pulsars with broad pulse profiles the variations in $P_2$ can be significant at large rotational phases. This follows from Equation (\ref{eq:P2InTime}), which implies that $P_2$ approaches infinity for $m\omega_{\rm dr} \rightarrow \omega_V$, and becomes negative for $m\omega_{\rm dr} < \omega_V$, implying a reverse in the sense of subpulse drift. Such reversal is predicted at large rotational phase, where $\omega_V$ increases towards its maximum at $\psi = \pm 180^\circ$. Depending on $\zeta$ and $\alpha$, there may exist a point in $\psi$ where $m\omega_{\rm dr} = \omega_V$ as shown in Figure \ref{fig-P2_2} at around $\psi\approx\pm 150^\circ$ and $\pm 170^\circ$ for $y= 0$ and 0.3, respectively.

\subsection{Frequency-dependent changes}
\label{sect:FrequencyDependent}

In the magnetic dipole model the angular width of the radio pulse profiles is frequency-dependent. Emission at lower frequencies, $\nu$, originates from larger heights, $r > r_{\rm min}$, and has a wider emission cone \citep{CC68, Cordes78, Thorsett91}. The spread in $\nu$ at a given $r$, however, is not known. 

Consider a highly idealized model in which (a) emission is only from the last closed field line, and (b) the emission from given $r$ is at a unique frequency, $\nu(r)$ say, which is a decreasing function of $r$. In this model, emission from the last closed field line implies $r=r_{\rm min} (\psi)$ (cf. Section \ref{sect:ModelVisibleEmission}), and observable emission is at $\nu=\nu(r_{\rm min}, \psi)$, which varies with $\theta_{bV} (\psi),\phi_{bV} (\psi)$ due to $r_{\rm min}$ being a unique function of these angles. In this model, the emission at the beginning and end of the pulse, that is, at $\psi=\psi_1,\psi_2$, is at a specific frequency, and within the pulse, the frequency increases as $\psi-\psi_1$ increases, reaching a maximum, and then decreasing as $\psi_2-\psi$ decreases. This implies a wider range of observable $\psi$ at lower frequencies as shown in Figures \ref{fig-TrajMeetNodes}, \ref{fig-P2_1} and \ref{fig-P2_3}. Because $P_2$ is a function of $\psi$, which increases as $|\psi|$ increases, the average $P_2$ over a broader profile also increases. This is consistent with observations, which show that $P_2$ is frequency-dependent, increasing as frequency decreases \citep{TMH75, SMK05}.

\subsection{Changes with rotation states}

Recent observations indicate that some radio pulsars can have multiple, quasi-stable configurations of their magnetospheres \citep{SLS00, SMK05, KLO+06, KSJ13}, and can switch abruptly between one configuration and another. In our model for non-corotating pulsar magnetospheres, such configurations are referred as rotation states and are differentiated by the value of the parameter $y$ \citep{MY14a}. An abrupt change in $y$ corresponds to an abrupt change in $\omega_{\rm dr}$, cf.\ Equations (\ref{eq:E}) and (\ref{eq:driftVel}), and hence to an abrupt change in $P_2$. 

For relatively large obliquity, $\alpha \gtrsim 45^\circ$, and small $\zeta$, $P_2$ is relatively insensitive to changes of rotation states for $\psi\ll 50^\circ$ but becomes significant for large $\psi$, where the value of $\omega_{\rm dr}$ may approach $\omega_V$. The magnitude of the change in $P_2$ between different $y$ at a given $\psi$ increases as $\alpha$ decreases. Assuming that $\alpha$ decreases with increasing age, this leads to the prediction that the effect of state switching on $P_2$ variation for a fixed $\psi$ or across a pulse window is larger for older pulsars. 

\subsection{A test case: PSR B0031-07}
\label{sect:TestCase}

We apply the foregoing model to PSR B0031-07, assuming $\zeta =6^\circ$ and $\alpha =5^\circ$ \citep{SMS+07}, modeling $P_2$ at three different emission heights for four different rotation states. The results are shown in Table \ref{table:P2_y}. The emission height is determined by the model described in Section \ref{sect:ModelVisibleEmission}. For a given $y \leq 0.6$, $P_2$ increases with $r$. Assuming a radius to frequency mapping, with lower frequencies from greater heights, this implies that $P_2$ increases as the observing frequency decreases \citep{TMH75}. The ratio $P_2/P_1$ becomes infinite and changes sign for large $y$. The model implies that if the pulsar were to switch from a value of $y$ below where the sign change occurs to a value above this change, then the sense of subpulse drifting would reverse. The frequency dependence also reverses such that $P_2$ decreases with $\nu$. Such reversal is not observed. One interpretation is that rotation states with large $y$ do not occur  in this pulsar.

Interpretation of the values in Table \ref{table:P2_y} depends on the drift rate and pulse window size. For example, consider the case $y = 0.6$. 

{\it Case 1:} Low drift rate and narrow pulse window, as in drift mode A at 4.85 GHz. This corresponds to $P_2/P_1$ = 3.05 and $r_{\rm min}^a = 2.6 \times 10^{-4}\, r_L$. In this case, a subpulse drifts sufficiently slowly such that it completes its traversal across the pulse window before the following subpulse appears. The driftbands are therefore non-overlapping in pulse period.  The vertical and horizontal separations at the start and finish of the two driftbands are three pulse periods and $\sim 15^\circ$, respectively.

{\it Case 2:} High drift rate and wide pulse window, as in drift mode A at 328 MHz. This corresponds to $P_2/P_1 = 3.10$ at $r_{\rm min}^c = 3.7 \times 10^{-4}\, r_L$. In this case, the subpulses drift sufficiently rapidly to give two `overlapping' driftbands. The vertical spacing between subpulses within a driftband is about six pulse periods. This spacing causes the emission from subpulses to be separated and the driftbands to appear `fainter' \citep{SMK05}. A smaller $y$ implies a smaller $P_2/P_1$, corresponding to a closer separation of subpulses in a driftband, cf. mode B of \cite{SMK05}. The vertical and horizontal separations between two subpulses (in two consecutive driftbands) are three pulse periods and $\sim 30^\circ$, respectively.

\begin{table}
\centering
\begin{tabular}{|l|ccc}
\hline
$y \backslash r_{\rm min}(\times 10^{-4} r_L)$ & $r_{\rm min}^a = 2.6$ & $r_{\rm min}^b = 3$ & $r_{\rm min}^c = 3.7$\\
\hline\hline
0 & 0.99 & 1.02 & 1.05\\ \hline
0.3 & 1.54 & 1.55 & 1.58\\ \hline
0.6 & 3.05 & 3.07 & 3.10\\ \hline
0.9 & --66.4 & --65.1 & --62.4\\ \hline
\hline
\end{tabular}
\caption{The value of $P_2/P_1$ is given for three emission heights, denoted $r_{\rm min}^a,r_{\rm min}^b,r_{\rm min}^c$ in units of $10^{-4}\,r_L$ (top row), for four rotation states of $y =0$, $0.3$, $0.6$ and $0.9$ (leftmost column). The $r_{\rm min}^a$ and $r_{\rm min}^c$ are determined exactly in our model based on the size of the pulse windows at 4.85 GHz (for $20^\circ$) and 328 MHz (for $30^\circ$), respectively, in drift-mode A \citep{SMK05}, and $r_{\rm min}^b$ is assumed to correspond to 1.41 GHz. The value of $P_2/P_1$ increases as $r$ increases for $y \leq 0.6$ and passes through infinity between $y=0.6$ and $y=0.9$, implying a reversal in the sense of subpulse drifting. The unique value of $P_2/P_1$ for each $y$ and $r$ can be used to derive the ratio of $P_2/P_1$ between different $r$ for a particular $y$.}
\label{table:P2_y}
\end{table}

\section{Summary and conclusions}
\label{sect:Conclusions}

In this paper, we point out an inconsistency in most treatments of subpulse drifting due to an incorrect assumption (that the line of sight is assumed fixed as the pulsar rotates). We investigate the consequence of the apparent motion of the visible point, which is assumed tangent to the magnetic field line in the source region. For a fixed line of sight the visible point would rotate at angular speed $\omega_*$ in a circle centered on the magnetic pole, whereas the visible point actually moves at non-constant angular speed around a non-circular trajectory that may or may not enclose the magnetic pole. We show how the separation between consecutive subpulses, $P_2$, is affected by taking the actual trajectory into account. We also explore the implications of one specific model for a non-corotating (oblique) pulsar magnetosphere. In this model the plasma drift velocity is described by a single parameter $0<y<1$, and a change in rotation state is attributed to a change in $y$. 

We refer to the difference between a model that includes the apparent motion of the visible point and a model that neglects it as a ``discrepancy''. For the specific model assumed in this paper, we quantify the discrepancy in terms of the parameter $R_d(y)$ introduced in Equation (\ref{eq:ratioParam}). The discrepancy is small for pulsars with large $\alpha$ and narrow pulse profiles, but becomes significant for pulsars with  small $\alpha$ and broad profiles.

Our model remains highly idealized, leading to limitations that need to be relaxed in a more detailed model for $P_2$. One limitation is that our model applies only to lowest order in an expansion in $r/r_L$. Various additional effects give corrections at different orders in this expansion, including non-dipolar fields, aberration, retardation and sweepback. To illustrate these effects, consider the particular case $y= 0$ (solid curve) in Figure \ref{fig-P2_3}, which has a pulse duty cycle of 0.1 centered at $\psi =0$. Then $P_2$ increases with increasing $|\psi|$, implying that neighboring anti-nodes appear to have a larger separation for large $\psi$ than for $\psi=0$. In this special case, the maximum change in $P_2$ across the pulse window is about $1\%$. Variations in $P_2$ have been observed with PSR B0809+74, which shows as an increase of about 1.5$\%$ from the leading to trailing parts across the profile \citep{Bartel81, WBS81, DLS+84}. The linear increase, as opposed to a symmetrical change, may be due to the fact that the pulse profile is asymmetric perhaps due to a non-dipolar field structure. Deviation from a dipolar field, $\propto1/r^3$, is always present in an oblique rotator, due to retardation implying an inductive term $\propto r^{-2}$ and the radiative term $\propto r^{-1}$ \citep{AE98, CRZ00}. Another possible modification to a centered dipolar field is an additional crust-anchored dipolar field \citep{GMM02}. Emission from such non-dipolar configurations can result in the center of the profile being shifted away from $\psi =0$. This would break the symmetry in the variations of $P_2$. Field-line distortion affects single-peak profiles, causing a difference in $P_2$ about a fiducial point ($\psi =0$ in our model). The sign of the difference signifies the rotation direction such that positive and negative values correspond to leftward and rightward rotations, respectively, relative to the line of sight.

In principle, it is possible to use data on $P_2$ to determine the rotation state, or at least to restrict the allowed range of $y$, as shown in Section \ref{sect:TestCase}. For example, in the case of PSR B0031-07 with duty cycle of about 0.1, a match in the predicted variations of $P_2$ across different $r$ with observations is found to correspond to $y= 0.6$. To achieve higher accuracy  requires precise knowledge of $\zeta$ and $\alpha$. For most pulsars, however, these two parameters are uncertain, limiting the reliability of any comparison between predictions and observations. A further uncertainty arises from an angular error introduced by the (incorrect) assumption that the line of sight passes through center of the star. This leads to an overestimation of $\alpha$, for $\beta>0$, or to an underestimation of $\alpha$, for $\beta<0$ \citep{YM14}.

The observable effects discussed in this paper are relatively small for most pulsars, especially for those with large $\alpha$ and narrow pulse profiles. Future availability of large arrays and telescopes, such as the SKA and FAST, will provide wide bandwidth coverage and unparallel sensitivity and time resolution, and will allow one to map the small changes in the subpulse driftbands predicted here. This will lead to a new era in investigating magnetospheric phenomena, significantly improve our understanding of drifting subpulses and of pulsar electrodynamics more generally.

\subsection{Summary}

We summarize our results in dot-point form.
\begin{itemize}
\item The path traced by the visible point on a sphere of radius $r$ in any pulsar magnetospheres is neither circular nor centered at the magnetic axis, except for special cases. 
\item Emission is assumed to be confined to anti-nodes which form spokes relative to the magnetic axis. A given spoke is  visible only if  trajectory of the visible point crosses it. 
\item Estimation from observation of the total number, $m$, of anti-nodes or spokes is possible only for a narrow pulse profile or $\zeta =0$.
\item Variations of $P_2$ result from the relative motion of drifting anti-nodes, at $\omega_{\rm dr}$, and the visible point, at $\omega_V$.
\item Abrupt changes of $P_2$ at a given $\psi$ are attributed to switching in the magnetospheres between different rotation states, each characterized by unique $y$ and associated $\omega_{\rm dr}(y)$.
\item $P_2$ is a function of $\psi$; it is a minimum at $\psi =0$ and increases as $\psi$ increases; the change in $P_2$ is small for $-50^\circ \leq\psi\leq 50^\circ$, and for pulsars with narrow pulse profiles.
\item Traditional models, which implicitly assume $\omega_V =0$, give good approximations for $P_2$ only for pulsars with $\alpha \gtrsim 45^\circ$, $m \gtrsim 30$ and narrow pulse profiles.
\end{itemize}

\section*{Acknowledgments} 

We thank Dick Manchester for useful comments, and V. Gajjar and H. Tong for helpful discussion. We also thank the anonymous referee for helpful suggestions which have improved the presentation of this paper. RY acknowledges supports from Project 11573059 NSFC; the Technology Foundation for Selected Overseas Chinese Scholar, Ministry of Personnel of China; the West Light Foundation of the Chinese Academy of Sciences project XBBS-2014-21; and the Strategic Priority Research Program ``The Emergence of Cosmological Structures" of the Chinese Academy of Sciences, Grant No. XDB09000000. Amirulhadi M. would also like to acknowledge the University of Malaya's HIR grant UM.S/625/3/HIR/28 for their funding. XH acknowledges supports from the Program of the Light in China's Western Region (LCRW) under grant nos. 2015-XBQN-B-03.

\bibliographystyle{mn2e}
%\bibliography{PulsarReferences} 

\begin{thebibliography}{}

\bibitem[\protect\citeauthoryear{Allafort, Baldini, Ballet, Barbiellini
  et~al.,}{Allafort et~al.}{2013}]{ABB+13}
Allafort A.,  Baldini L.,  Ballet J.,  Barbiellini G.,    et~al., 2013, ApJL,
  777, L2

\bibitem[\protect\citeauthoryear{Arendt \& Eilek}{Arendt \& Eilek}{1998}]{AE98}
Arendt P.~N.,  Eilek J.~A.,  1998, arXiv:astro-ph/9801257v1

\bibitem[\protect\citeauthoryear{Arons}{Arons}{1981}]{Arons1981}
Arons J.,  1981, ApJ, 248, 1099

\bibitem[\protect\citeauthoryear{Backer}{Backer}{1973}]{Backer73}
Backer D.~C.,  1973, ApJ, 182, 245

\bibitem[\protect\citeauthoryear{Bartel}{Bartel}{1981}]{Bartel81}
Bartel N.,  1981, Astron. Astrophys., 97, 384

\bibitem[\protect\citeauthoryear{Bhattacharyya, Gupta \& Gil}{Bhattacharyya
  et~al.}{2009}]{BGG09}
Bhattacharyya B.,  Gupta Y.,    Gil J.,  2009, MNRAS, 398, 1435

\bibitem[\protect\citeauthoryear{Cheng, Ruderman \& Zhang}{Cheng
  et~al.}{2000}]{CRZ00}
Cheng A.~F.,  Ruderman M.~A.,    Zhang L.,  2000, ApJ, 537, 964

\bibitem[\protect\citeauthoryear{Clemens \& Rosen}{Clemens \&
  Rosen}{2004}]{CR04}
Clemens J.~C.,  Rosen R.,  2004, ApJ, 609, 340

\bibitem[\protect\citeauthoryear{Cordes}{Cordes}{1978}]{Cordes78}
Cordes J.~M.,  1978, ApJ, 222, 1006

\bibitem[\protect\citeauthoryear{Craft \& Comella}{Craft \&
  Comella}{1968}]{CC68}
Craft H.~D.,  Comella J.~M.,  1968, Nature, 220, 676

\bibitem[\protect\citeauthoryear{Davies, Lyne, Smith, Izvekova, Kuzmin \&
  Shitov}{Davies et~al.}{1984}]{DLS+84}
Davies J.~G.,  Lyne A.~G.,  Smith F.~G.,  Izvekova V.~A.,  Kuzmin A.~D.,
  Shitov Y.~P.,  1984, MNRAS, 211, 57

\bibitem[\protect\citeauthoryear{Deshpande \& Rankin}{Deshpande \&
  Rankin}{1999}]{DR99}
Deshpande A.~A.,  Rankin J.~M.,  1999, ApJ, 524, 1008

\bibitem[\protect\citeauthoryear{Dyks, Rudak \& Harding}{Dyks
  et~al.}{2004}]{DRH04}
Dyks J.,  Rudak B.,    Harding A.~K.,  2004, ApJ, 607, 939

\bibitem[\protect\citeauthoryear{Edwards \& Stappers}{Edwards \&
  Stappers}{2003}]{ES03}
Edwards R.~T.,  Stappers B.,  2003, A\&A, 407, 273

\bibitem[\protect\citeauthoryear{Fung, Khechinashvili \& Kuijpers}{Fung
  et~al.}{2006}]{FKK06}
Fung P.~K.,  Khechinashvili D.,    Kuijpers J.,  2006, A\&A, 445, 779

\bibitem[\protect\citeauthoryear{Gangadhara}{Gangadhara}{2004}]{Gangadhara04}
Gangadhara R.~T.,  2004, ApJ, 609, 335

\bibitem[\protect\citeauthoryear{Gangadhara \& Gupta}{Gangadhara \&
  Gupta}{2001}]{GG01}
Gangadhara R.~T.,  Gupta Y.,  2001, ApJ, 555, 31

\bibitem[\protect\citeauthoryear{Gil, Melikidze \& Mitra}{Gil
  et~al.}{2002}]{GMM02}
Gil J.~A.,  Melikidze G.~I.,    Mitra D.,  2002, A\&A, 388, 235

\bibitem[\protect\citeauthoryear{Godoberidze, Machabeli, Melrose \&
  Luo}{Godoberidze et~al.}{2005}]{GMM+05}
Godoberidze G.,  Machabeli G.~Z.,  Melrose D.~B.,    Luo Q.,  2005, MNRAS, 360,
  669

\bibitem[\protect\citeauthoryear{Hassall, Stappers, Weltevrede, Hessels,
  Alexov, Coenen, Karastergiou \& Kramer}{Hassall et~al.}{2013}]{HSW+13}
Hassall T.~E.,  Stappers B.~W.,  Weltevrede P.,  Hessels J. W.~T.,  Alexov A.,
  Coenen T.,  Karastergiou A.,    Kramer M. et.~al.,  2013, A\&A, 552, A61

\bibitem[\protect\citeauthoryear{Jones}{Jones}{1984}]{Jones1984}
Jones P.~B.,  1984, MNRAS, 209, 569

\bibitem[\protect\citeauthoryear{Karastergiou \& Johnston}{Karastergiou \&
  Johnston}{2007}]{KJ07}
Karastergiou A.,  Johnston S.,  2007, MNRAS, 380, 1678

\bibitem[\protect\citeauthoryear{Kazbegi, Machabeli \& Melikidze}{Kazbegi
  et~al.}{1991}]{KMM91a}
Kazbegi A.~Z.,  Machabeli G.~Z.,    Melikidze G.~I.,  1991, AuJPh, 44, 573

\bibitem[\protect\citeauthoryear{Keith, Shannon \& Johnston}{Keith
  et~al.}{2013}]{KSJ13}
Keith M.~J.,  Shannon R.~M.,    Johnston S.,  2013, MNRAS, 432, 3080

\bibitem[\protect\citeauthoryear{Kijak \& Gil}{Kijak \& Gil}{2003}]{KG03}
Kijak J.,  Gil J.,  2003, A\&A, 397, 969

\bibitem[\protect\citeauthoryear{Kramer, Lyne, O'Brien, Jordan \&
  Lorimer}{Kramer et~al.}{2006}]{KLO+06}
Kramer M.,  Lyne A.~G.,  O'Brien J.~T.,  Jordan C.~A.,    Lorimer D.~R.,  2006,
  Science, 312, 549

\bibitem[\protect\citeauthoryear{Lyne, Hobbs, Kramer, Stairs \& Stappers}{Lyne
  et~al.}{2010}]{LHK+10}
Lyne A.~G.,  Hobbs G.,  Kramer M.,  Stairs I.~H.,    Stappers B.,  2010,
  Science, 329, 408

\bibitem[\protect\citeauthoryear{Melrose \& Yuen}{Melrose \&
  Yuen}{2014}]{MY14a}
Melrose D.~B.,  Yuen R.,  2014, MNRAS, 437, 262

\bibitem[\protect\citeauthoryear{Mitra \& Rankin}{Mitra \& Rankin}{2008}]{MR08}
Mitra D.,  Rankin J.~M.,  2008, MNRAS, 385, 606

\bibitem[\protect\citeauthoryear{Morozova, Ahmedov \& Zanotti}{Morozova
  et~al.}{2014}]{MAZ14}
Morozova V.~S.,  Ahmedov B.~J.,    Zanotti O.,  2014, MNRAS, 444, 1144

\bibitem[\protect\citeauthoryear{Radhakrishnan \& Cooke}{Radhakrishnan \&
  Cooke}{1969}]{RC69_RVM}
Radhakrishnan V.,  Cooke D.~J.,  1969, Astrophys. Lett., 3, 225

\bibitem[\protect\citeauthoryear{Ruderman \& Sutherland}{Ruderman \&
  Sutherland}{1975}]{RudermanSutherland1975}
Ruderman M.,  Sutherland P.~G.,  1975, ApJ, 196, 51

\bibitem[\protect\citeauthoryear{Serylak, Stappers \& Weltevrede}{Serylak
  et~al.}{2009}]{SSW09}
Serylak M.,  Stappers B.~W.,    Weltevrede P.,  2009, A\&A, 506, 865

\bibitem[\protect\citeauthoryear{Smits, Mitra \& Kuijpers}{Smits
  et~al.}{2005}]{SMK05}
Smits J.~M.,  Mitra D.,    Kuijpers J.,  2005, A\&A, 440, 683

\bibitem[\protect\citeauthoryear{Smits, Mitra, Stappers, Kuijpers, Weltevrede,
  Jessner \& Gupta}{Smits et~al.}{2007}]{SMS+07}
Smits J.~M.,  Mitra D.,  Stappers B.~W.,  Kuijpers J.,  Weltevrede P.,  Jessner
  A.,    Gupta Y.,  2007, A\&A, 465, 575

\bibitem[\protect\citeauthoryear{Stairs, Lyne \& Shemar}{Stairs
  et~al.}{2000}]{SLS00}
Stairs I.~H.,  Lyne A.~G.,    Shemar S.~L.,  2000, Nature, 406, 484

\bibitem[\protect\citeauthoryear{Sutton, Stelin, Price \& Weimer}{Sutton
  et~al.}{1970}]{SSP+70}
Sutton J.~M.,  Stelin D.~H.,  Price R.~M.,    Weimer R.,  1970, ApJ, 159, L89

\bibitem[\protect\citeauthoryear{Taylor, Manchester \& Huguenin}{Taylor
  et~al.}{1975}]{TMH75}
Taylor J.~H.,  Manchester R.~N.,    Huguenin G.~R.,  1975, ApJ, 195, 513

\bibitem[\protect\citeauthoryear{Thorsett}{Thorsett}{1991}]{Thorsett91}
Thorsett S.~E.,  1991, ApJ, 377, 263

\bibitem[\protect\citeauthoryear{van Leeuwen \& Timokhin}{van Leeuwen \&
  Timokhin}{2012}]{LT12}
van Leeuwen A. G.~J.,  Timokhin A.~N.,  2012, ApJ, 752, 155

\bibitem[\protect\citeauthoryear{Wolszczan, Bartel \& Sieber}{Wolszczan
  et~al.}{1981}]{WBS81}
Wolszczan A.,  Bartel N.,    Sieber W.,  1981, Astron. Astrophys., 100, 91

\bibitem[\protect\citeauthoryear{Yuen \& Melrose}{Yuen \& Melrose}{2014}]{YM14}
Yuen R.,  Melrose D.~B.,  2014, Publ. Astron. Soc. Aust. (PASA), 31, e039

\end{thebibliography}

\appendix
\section{Coordinate Transformations}
\label{sec:CoordTansf}

The transformation between the Cartesian vectors in the magnetic frame (subscript $b$) and the non-rotating frame fixed relative to the center of the star is
\be
\left(
\begin{array}{c}
\hat{{\bf x}}_b\\{\hat{\bf y}}_b\\{\hat{\bf z}}_b
\end{array}
\right)=
[{\bf R}]
\left(
\begin{array}{c}
{\hat{\bf x}}\\{\hat{\bf y}}\\{\hat{\bf z}}
\end{array}
\right),
\label{transf1v}
\ee
with the transformation matrix given by
\be
[{\bf R}]=\left(
\begin{array}{ccc}
\cos\alpha\cos\psi&\cos\alpha\sin\psi&-\sin\alpha\\
-\sin\psi&\cos\psi&0\\
\sin\alpha\cos\psi&\sin\alpha\sin\psi&\cos\alpha
\end{array}
\right).
\label{transf2}
\ee
The inverse transform is
\be
\left(
\begin{array}{c}
\hat{{\bf x}}\\{\hat{\bf y}}\\{\hat{\bf z}}
\end{array}
\right)=
[{\bf R}]^T
\left(
\begin{array}{c}
\hat{{\bf x}}_b\\{\hat{\bf y}}_b\\{\hat{\bf z}}_b
\end{array}
\right),
\label{transf3v}
\ee
with $[{\bf R}]^T$ the transpose of the matrix $[{\bf R}]$ given by (\ref{transf2}).

The transformation between Cartesian and spherical polar vectors is given by 
\be
\left(
\begin{array}{c}
\hat{{\bf r}}\\{\hat\btheta}\\{\hat\bphi}
\end{array}
\right)=
[{\bf P}]
\left(
\begin{array}{c}
\hat{{\bf x}}\\{\hat{\bf y}}\\{\hat{\bf z}}
\end{array}
\right),
\label{transf6}
\ee
with
\be
[{\bf P}]=\left(
\begin{array}{ccc}
\sin\theta\cos\phi&\sin\theta\sin\phi&\cos\theta\\
\cos\theta\cos\phi&\cos\theta\sin\phi&-\sin\theta\\
-\sin\phi&\cos\phi&0
\end{array}
\right).
\label{transf7}
\ee
The inverse transform corresponding to (\ref{transf6}) has a transformation matrix that is the transpose of (\ref{transf7}). The corresponding transformation for the vectors relative to the magnetic axis follows from (\ref{transf6}) and (\ref{transf7}) by adding subscripts $b$.

Relations between angles follow from (\ref{transf1v}) by projecting onto the position vector, giving
\be
\left(
\begin{array}{c}
\sin\theta_b\cos\phi_b\\
\sin\theta_b\sin\phi_b\\
\cos\theta_b
\end{array}
\right)=
[{\bf R}]
\left(
\begin{array}{c}
\sin\theta\cos\phi\\
\sin\theta\sin\phi\\
\cos\theta
\end{array}
\right).
\label{transf1}
\ee
Similarly, the inverse transformation between the angles follows from (\ref{transf3v}).
\end{document}